\documentclass[11pt]{article}
\usepackage{moriond,epsfig,latexsym,amsmath,amssymb,theorem,graphicx}

\def\be{\begin{equation}}
\def\ee{\end{equation}}
\def\bea{\begin{eqnarray}}
\def\eea{\end{eqnarray}}

\newcommand{\kk}{{\bf k}}

\newcommand{\es}{\epsilon_s}
\newcommand{\eps}{\epsilon_s}
\newcommand{\ep}{\epsilon}
\newcommand{\cbt}{\bar{c}_s^2}

\newcommand{\fnl}{f_{\rm NL}}

\def\gsim{ \lower .75ex \hbox{$\sim$} \llap{\raise .27ex \hbox{$>$}} }
\def\lsim{ \lower .75ex \hbox{$\sim$} \llap{\raise .27ex \hbox{$<$}} }
\def\be{\begin{equation}}
\def\ee{\end{equation}}
\def\bea{\begin{eqnarray}}
\def\eea{\end{eqnarray}}

\renewcommand{\d}{\mathrm{d}}

\usepackage{graphicx}

\newcommand{\ba}{\begin{array}}
\newcommand{\ea}{\end{array}}

\newcommand{\comment}[1]{}
\begin{document}
\vspace*{4cm}
\title{CONSTRAINING FAST-ROLL INFLATION}

\author{JOHANNES NOLLER}

\address{Theoretical Physics, Blackett Laboratory, Imperial College, London, SW7 2BZ, UK}

\maketitle\abstracts{We present constraints on how far single field inflation may depart from the familiar slow-roll paradigm. Considering a fast-roll regime while requiring a (near)-scale-invariant power spectrum introduces large self-interactions for the field and consequently large and scale-dependent non-Gaussianities. Employing this signal, we use the requirement of weak-coupling together with WMAP constraints to derive bounds on generic $P(X,\phi)$ theories of single field inflation.
}


\noindent {\bf Introduction:} Recently much progress has been made in understanding inflationary phenomenology beyond the slow-roll paradigm~\cite{kp,nm,burrage,Ribeiro:2011ax,bg,rr}, i.e. where inflation is not almost de Sitter. An important question then poses itself: How far may we depart from the standard slow-roll regime without coming into conflict with observational and theoretical constraints? More specifically, what bounds can we place on ``slow roll'' parameters (which measure the ``distance'' from purely de Sitter expansion)? Here we present a number of such constraints for generic classes of inflationary single field models.

Departure from pure de Sitter expansion generically breaks the scale invariance of n-point correlation functions for the curvature perturbation $\zeta$. However, present-day data constrain the 2-point function (the power spectrum) to be near scale-invariant~\cite{wmapcosmo}. Generic single field models can restore this observed behaviour via the introduction of non-canonical kinetic terms and hence a time-varying ``speed of sound'' $c_s$. In doing so, large interaction terms are produced at the level of the cubic action, leading to the generation of large levels of non-Gaussianity. These will be heavily constrained by CMB and large scale structure surveys in the near future and as such non-Gaussianity becomes an excellent tool for constraining slow-roll parameters.

\vspace{.3cm}

\noindent {\bf The setup:} We consider general single field inflation models described by an action
\begin{equation} \label{PX}
S=\int {\rm d}^4x \sqrt{-g} \left[\frac{R}{2} +
P(X,\phi)\right]~, 
\end{equation}
where $X=-\frac{1}{2}g^{\mu\nu}\partial_{\mu}\phi \partial_{\nu}\phi$. It is useful to introduce a hierarchy of slow-roll parameters
\begin{eqnarray}
\epsilon \equiv - \frac{\dot H}{H^2},\;
\eta \equiv \frac{\dot \epsilon}{\epsilon H},... \qquad
\es \equiv \frac{\dot c_s}{c_s H},\;
\eta_s \equiv \frac{\dot \es}{\es H},... \qquad,\text{where} \; c_s^2 = \frac{P_{,X}}{P_{,X}+2X P_{,XX}} \label{slowroll}
\end{eqnarray}
where $a$ is the scale factor of an FRW metric and $H(t) = {\dot a}/a$ is the corresponding Hubble rate. $c_s$ is the speed of sound with which perturbations propagate, essentially quantifying the non-canonical nature of~\eqref{PX}. A near de Sitter expansion is associated with the slow-roll regime, where $\ep,\eta ... \eps, \eta_s... \ll 1$ and accelerated expansion takes place as long as $\ep < 1$. We now wish to understand what constraints can be placed on these parameters. For simplicity we here focus on the case where slow-roll is broken at the first level in the hierarchy, for $\epsilon$ and $\epsilon_s$, but assume that slow-roll conditions still hold for higher order parameters, setting $\eta,\eta_s \sim 0$. 

In deriving the constraints presented here we will firstly map present-day observational bounds, especially those coming from the WMAP experiment~\cite{wmapcosmo}, onto the parameter space of fast-rolling models. As a second guidance principle we will impose a minimal theoretical constraint: 
The fluctuations described by~\eqref{PX} should remain weakly coupled for at least $\sim 10$ e-folds. This range corresponds to the observable window of scales where primordial non-Gaussianity may be measured (running from CMB, $k^{-1} \sim 10^3$ Mpc, to galactic scales, $k^{-1} \sim 1$ Mpc). 

Why require weak coupling at all? Strong coupling scales are frequently associated with the appearance of new physics. In the standard model, for example, the would-be strong coupling scale lies around $\sim 1 \; TeV$, before the Higgs is introduced. We may expect an analogue to be true for single field inflation models, especially given the generic presence of other massive degrees of freedom (dof) in UV completions of primordial physics. Such {\it dofs} may be integrated out at low energies, but can become relevant around the would-be strong coupling scale~\cite{gelaton,baumanngreen}. If so, predictions beyond this scale will depend on exactly how and which {\it dofs} enter. Flipping the argument around, even if we were able to calculate the dynamics for generic strongly coupled systems, one should remain cautious whether the effective field theory under consideration is valid anymore in such circumstances. As such we will require weak coupling to ensure that~\eqref{PX} is predictive over at least the observable window of scales where primordial fluctuations may be measured.  


\begin{figure}[t]
\begin{center}$
\begin{array}{ccc}
\includegraphics[width=0.3\linewidth]{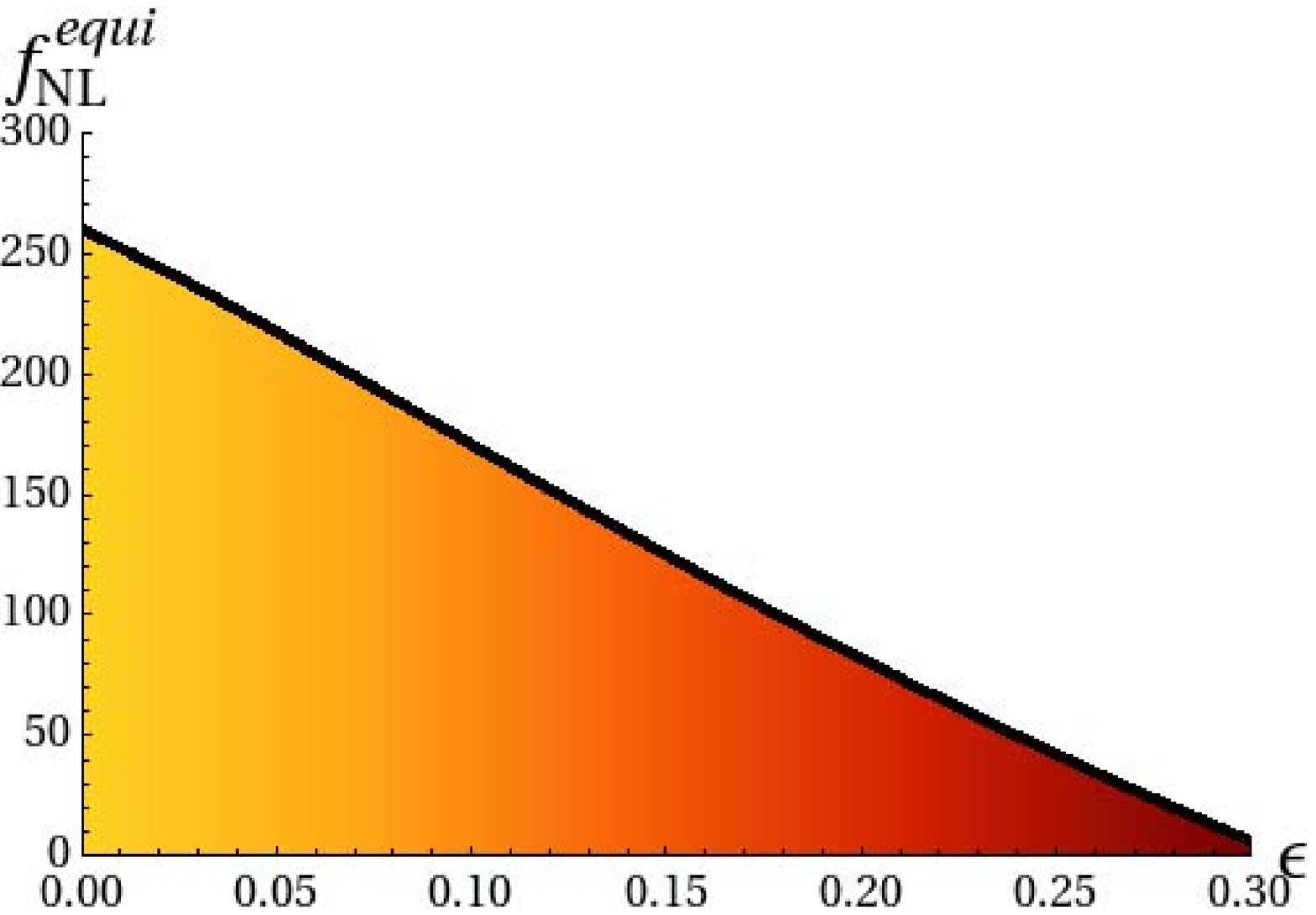} &
\includegraphics[width=0.3\linewidth]{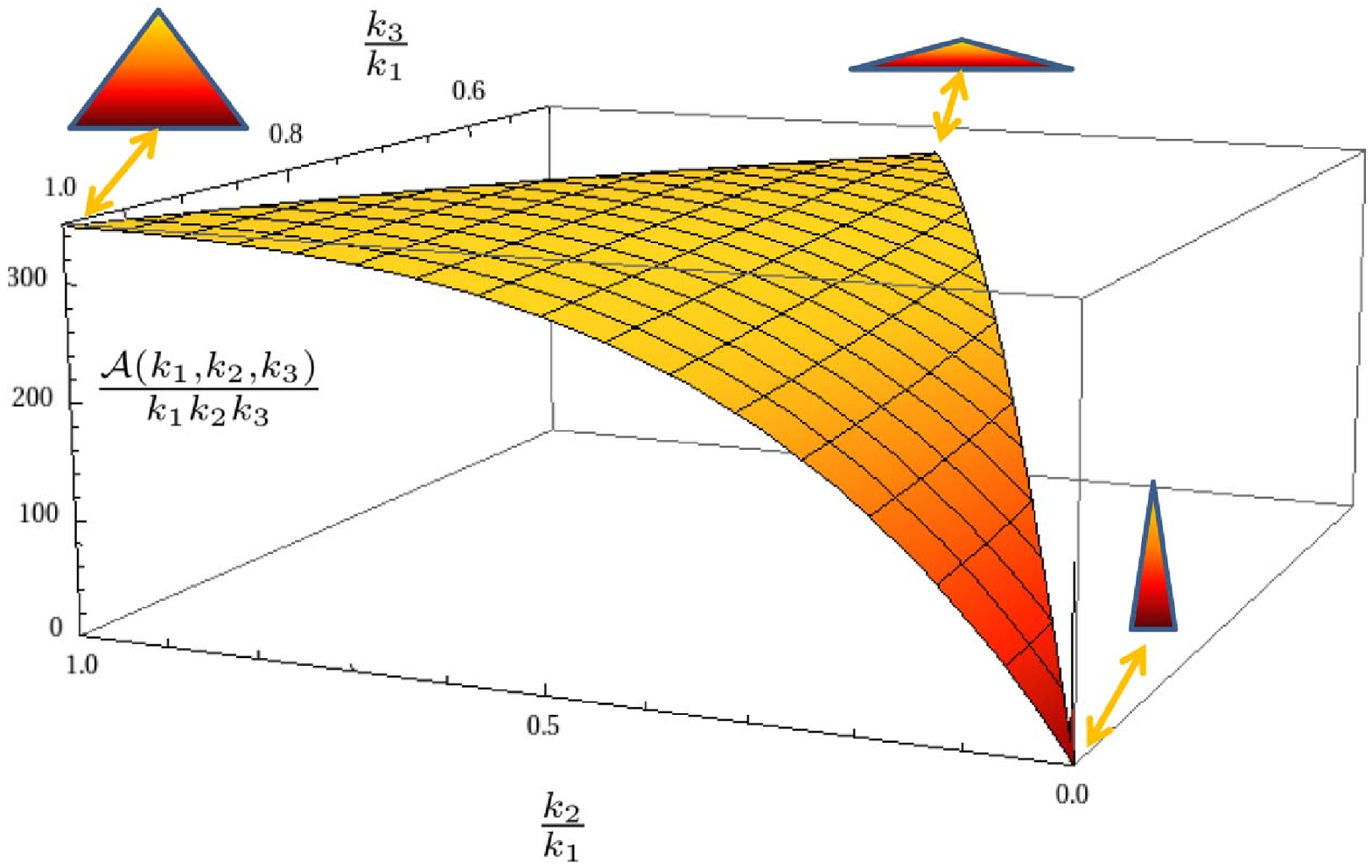} &
\includegraphics[width=0.3\linewidth]{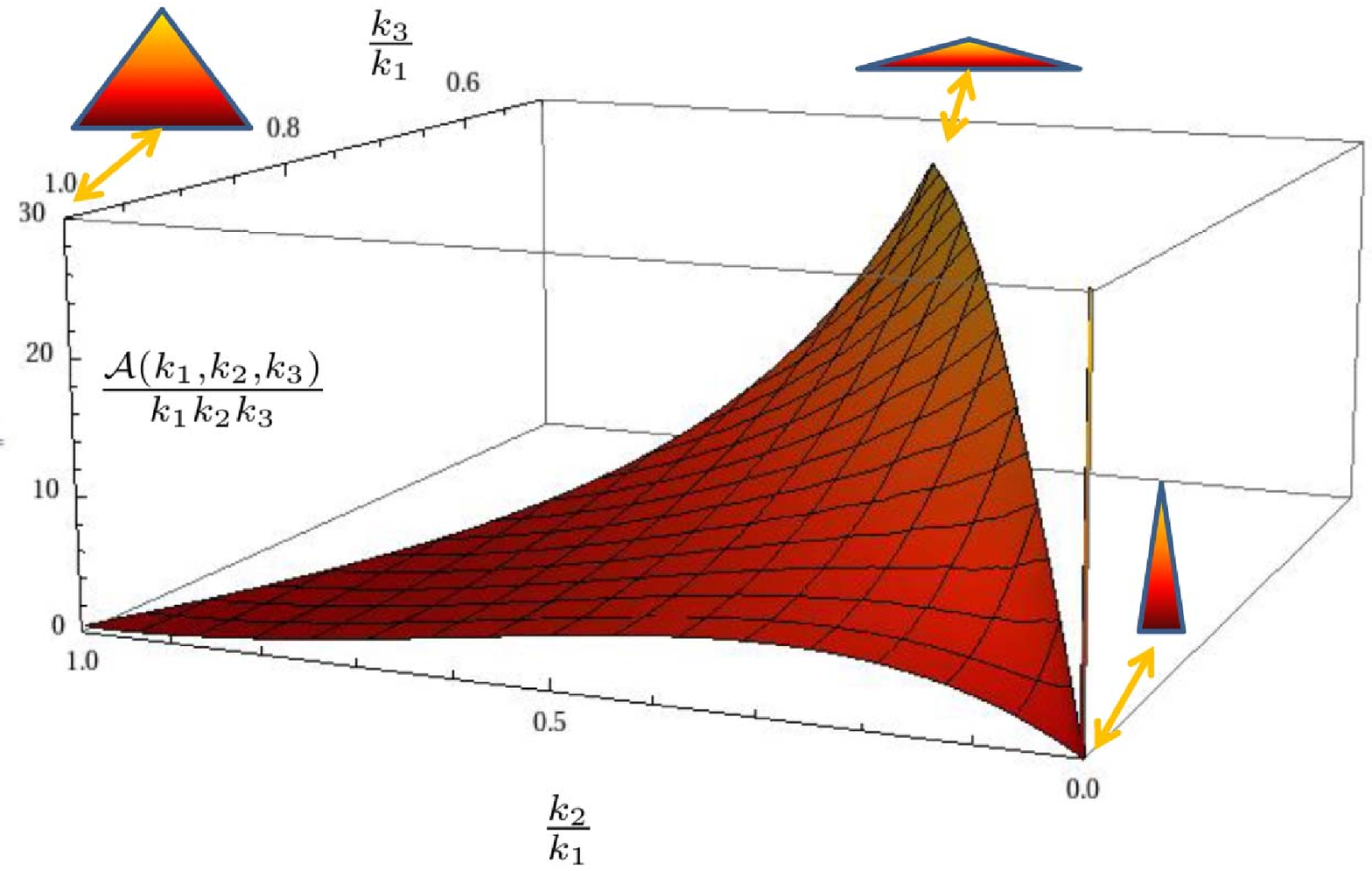}
\end{array}$
\end{center}
\caption{
\footnotesize{{\bf Left}: $\fnl^{equi}$ vs. $\epsilon$ for a horizon-crossing $\bar{c}_s = 0.1$ and $f_X = -70$.
{\bf Middle}: The dimensionless bispectrum ${\cal A}(k_1,k_2,k_3)/(k_1 k_2 k_3)$ plotted in the slow-roll limit $\epsilon \to 0$ for $\bar{c}_s = 0.07$ and $f_X = -53$. Triangular shapes denote the equilateral, enfolded and squeezed/local limit clockwise from top left.
{\bf Right}: Analogous plot for $\epsilon = 0.3$. Note how the overall amplitude is suppressed and the shape has changed, now peaking in the enfolded limit.
}}
\label{fig1}
\end{figure}

\vspace{.3cm}

\noindent {\bf Non-Gaussian signals:} The 2-point function of the curvature perturbation $\zeta$ is given by
\begin{equation} \label{2point}
\langle \zeta(\textbf{k}_1)\zeta(\textbf{k}_2)\rangle = (2\pi)^5
\delta^3(\kk_1+\kk_2)  \frac{P_\zeta}{2 k_1^3}\qquad, \qquad n_s - 1 \equiv \frac{d \text{ln} P_\zeta}{d \text{ln} k} = \frac{2 \epsilon + \es}{\es + \epsilon -1}\;,
\end{equation}
where $P_\zeta$ is the power spectrum. Taking a lead from~\cite{kp} we will focus on computing non-Gaussian signals assuming an exactly scale-invariant 2-point function $n_s = 1$ in what follows, requiring $\es = -2\epsilon$. For $(n_s-1)$-dependent (and hence suppressed) corrections see~\cite{nm,burrage,rr}. The 3-point function measures the strength of interactions of the field\footnote{A free field is Gaussian, hence the 3-point function captures the non-Gaussian statistics for $\zeta$.} which are described by the interaction vertices in the cubic action~\cite{seerylidsey,burrage}
\begin{equation}
		S_3
		=
		\int \d^3 x \, \d \tau \; a^2
		\bigg\{
			\frac{\Lambda_1}{a} \zeta'^3
			+ \Lambda_2 \zeta \zeta'^2
			+ \Lambda_3 \zeta (\partial \zeta)^2
			+ \Lambda_4 \zeta' \partial_j \zeta \partial_j \partial^{-2} \zeta'
			+ \Lambda_5 \partial^2 \zeta
			(\partial_j \partial^{-2} \zeta')
			(\partial_j \partial^{-2} \zeta')
		\bigg\} .
\end{equation}
The 3-point function itself can be expressed through the amplitude ${\cal A}$
\begin{equation} \label{amplidef}
\langle \zeta(\textbf{k}_1)\zeta(\textbf{k}_2)\zeta(\textbf{k}_3)\rangle = (2\pi)^7
\delta^3(\kk_1+\kk_2+\kk_3) P_\zeta^{\;2} \frac{1}{\Pi_j k_j^3}{\cal A}\, \qquad, \qquad f_{\rm NL}^{equi} = 30\frac{{\cal A}_{k_1=k_2=k_3}}{K^3},
\end{equation}
where $f_{\rm NL}^{equi}$ serves as a convenient single number measure of the amplitude of non-Gaussianity in the equilateral limit $k_1=k_2=k_3$ and the power spectrum $P_\zeta$ is conventionally calculated for the mode $K = k_1 + k_2 + k_3$. The size of non-Gaussianities is then a function of the parameters $\left\{ c_s^{-2},f_X,\epsilon,n_s \right\}$, where $f_X$\footnote{We concentrate on solutions for which $f_X$ is constant here. Also note that for DBI models the first interaction vertex vanishes $\Lambda_1 = 0$.}  essentially measures the strength of the first interaction vertex $\zeta'^3$ and satisfies~\cite{burrage}
\be
\Lambda_1 = \frac{2 \epsilon}{3 H c_s^4}\left( 1 - c_s^2 - f_X\right).
\ee          
$\fnl^{equi}$ can then be estimated by $\fnl^{equi} \sim {\cal O}(c_s^{-2}) + {\cal O}(\frac{f_X}{c_s^2})$. Since requiring a scale-invariant 2-point function yields $\es = -2\epsilon$, fast-roll models with $\epsilon \sim {\cal O}(1)$ lead to a rapidly decreasing $c_s$ (as long as inflation is not ghost-like, i.e. $\epsilon \not< 0$). They therefore naturally yield regimes where $c_s$ is small and the 3-point function is large. A useful fitting formula in this context is~\cite{kp}
\be
f_{\rm NL}^{equi} = 0.27 - \frac{0.164}{\cbt} - (0.12 + 0.04 f_X)\frac{1}{\cbt}\left({1 - \frac{4\epsilon}{1 + \epsilon}}\right). \label{fitting}
\ee
This has three significant consequences for fast-roll phenomenology:
\begin{figure}[t] 
\begin{center}$
\begin{array}{ccc}
\includegraphics[width=0.34\linewidth]{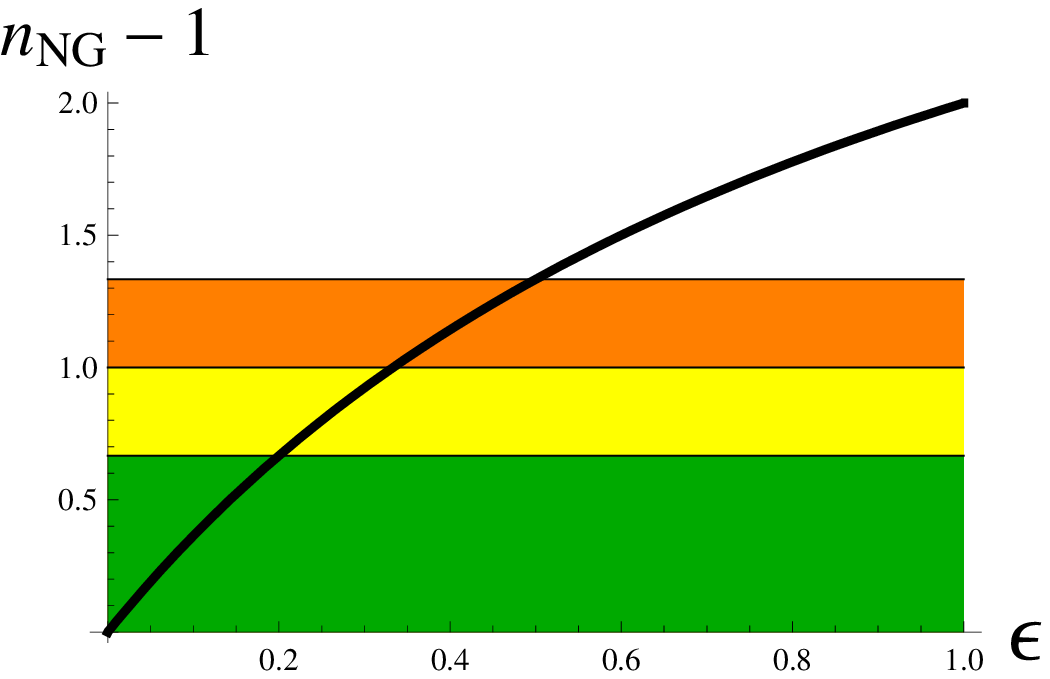} &
\includegraphics[width=0.29\linewidth]{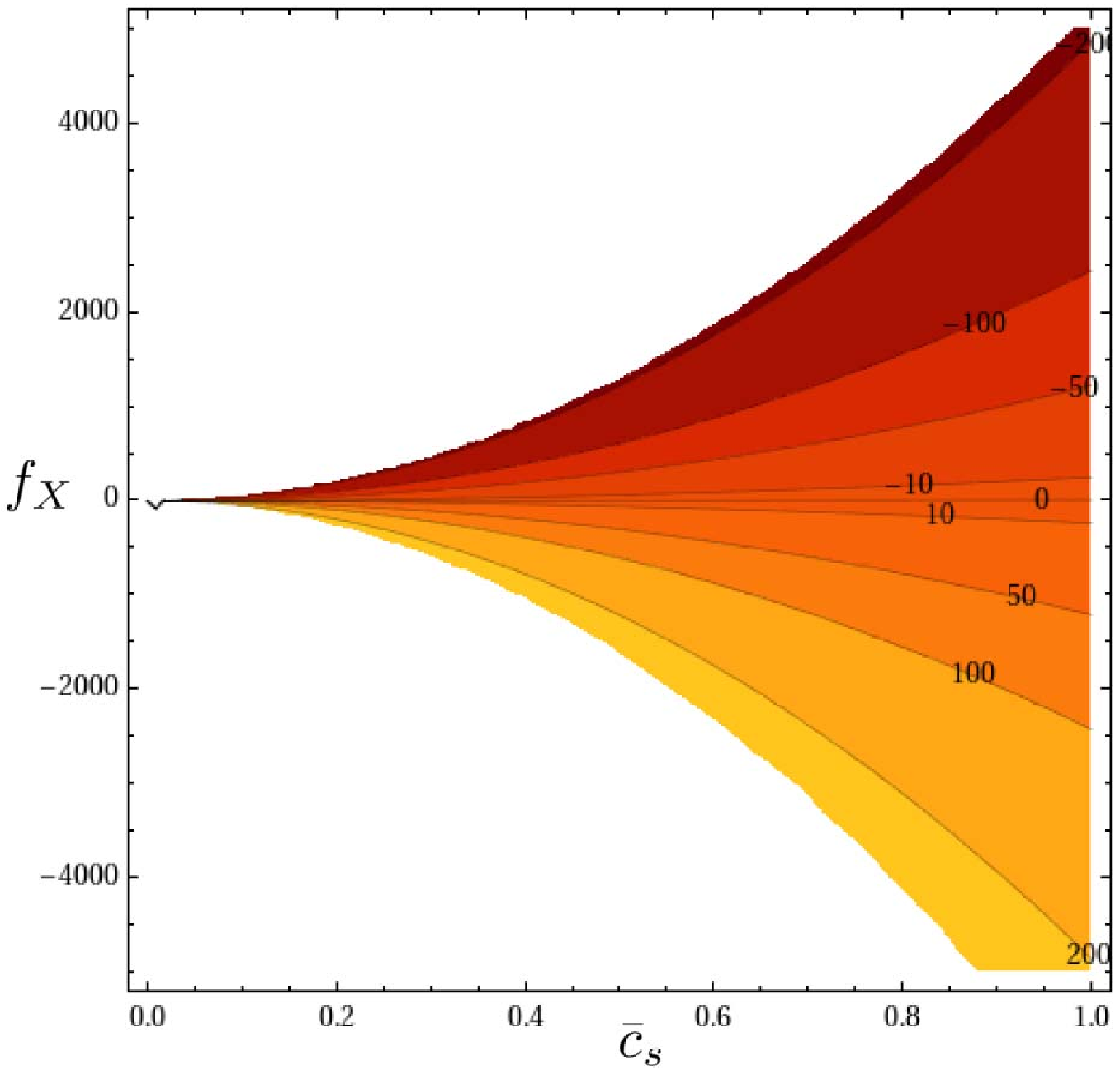} &
\includegraphics[width=0.29\linewidth]{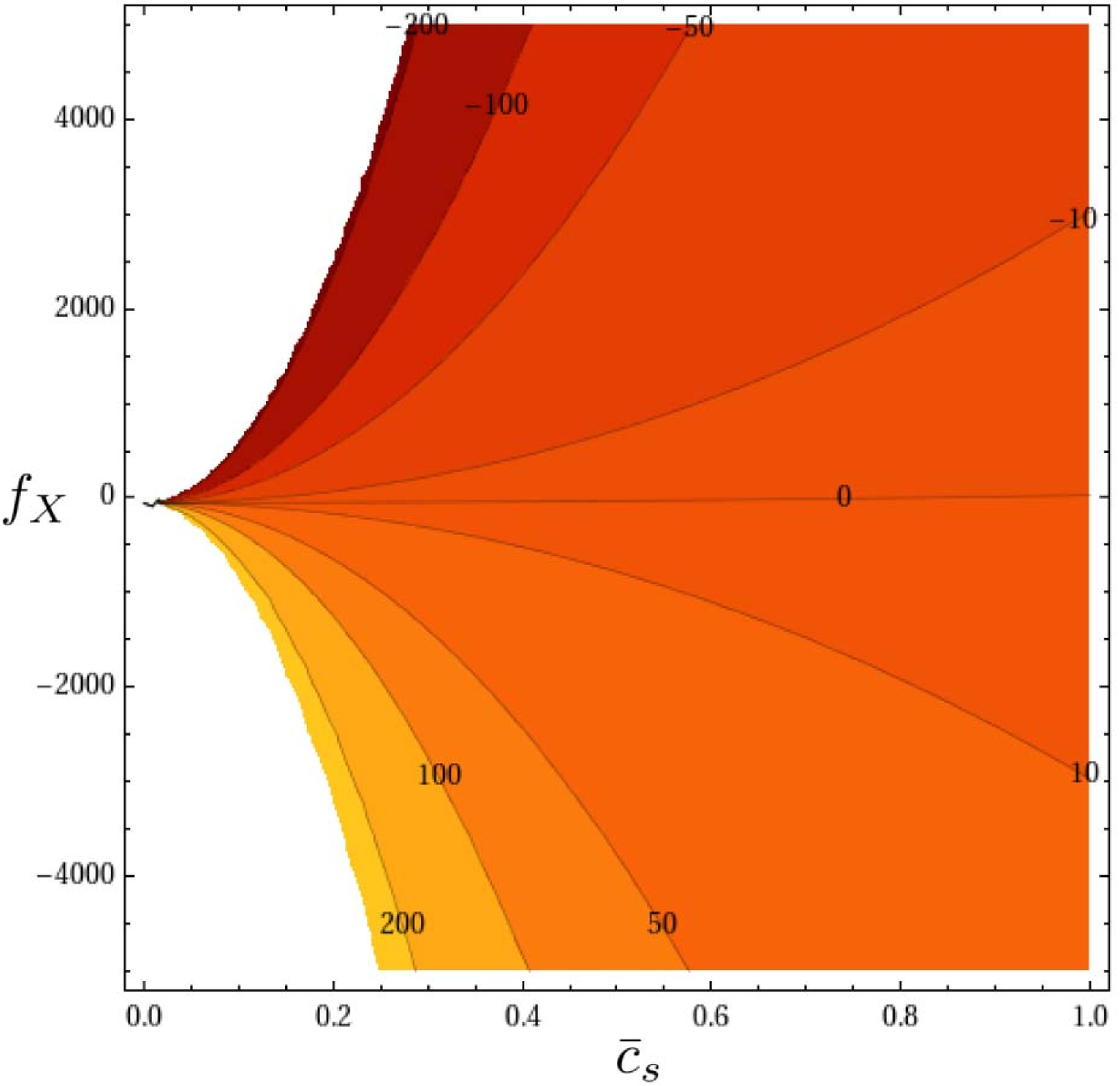}
\end{array}$
\end{center}
\caption{
\footnotesize{{\bf Left}: $n_{\rm NG} - 1$ plotted against $\ep$ in the small $c_s$ limit. Green, yellow and orange ($< 2/3,<1,<4/3$) regions are those allowed by perturbative constraints assuming $f_{\rm NL}^{equi}({\rm CMB}) \sim {\cal O}(100),{\cal O}(10),{\cal O}(1)$ respectively.
{\bf Middle}: Contour plot showing the region in parameter space allowed by the WMAP $2\sigma$ constraint $f_{\rm NL}^{\rm equil} = 26 \pm 240$ in the slow-roll limit $\epsilon \to 0$.
{\bf Right}: Analogous plot for $\epsilon = 0.3$. Note how the allowed region widens.
}}
\label{fig2}
\end{figure}
\begin{itemize}
\item {\bf Fast-roll suppresses $\fnl$.} Figure~\ref{fig1} and eqn.~\eqref{fitting} show that for equilateral non-Gaussianity, $\fnl^{equi}$ is generically reduced when departing from the slow-roll regime. This implies that models with e.g. such a small speed of sound $c_s$ that they violate observational constraints in the slow-roll limit, can still be allowed when considering fast-roll scenarios.
\item {\bf The shape of the bispectrum is modified.} Figure~\ref{fig1} also shows that fast-roll suppression is not an artefact of focusing on the equilateral limit, but that in fact the full bispectrum as described by ${\cal A}$ is fast-roll suppressed. Furthermore the shape of the amplitude is modified - full details are given in~\cite{kp,nm}, but figure~\ref{fig1} illustrates a particular case where a predominantly equilateral shape is altered into an ``enfolded'' shape, peaking in the limit $2k_1=k_2=k_3$. 
\item {\bf The allowed parameter-space for $f_X,c_s$ becomes wider.} As a result of fast-roll suppression observational bounds, e.g. the WMAP result~\cite{wmapcosmo} $f_{\rm NL}^{\rm equil} = 26 \pm 240$ at $95\%$ confidence, map onto weaker constraints for parameters $f_X,c_s$ at the expense of enlarged $\epsilon$. Figure~\ref{fig2} shows how constraints are altered, cf.~\cite{senatore}.  
\end{itemize}

\noindent {\bf Induced blue running of non-Gaussianities $\to$ strong coupling constraints:} The dependence of the 3-point function on $K$ can be described by the parameter $n_{\rm NG} - 1 \equiv {\rm d}\ln |f_{\rm NL}^{equi}|/{\rm d}\ln K$. Expanding around the phenomenologically motivated small $c_s$ limit, we find
\be \label{nng}
n_{\rm NG} - 1 = \frac{4 \ep}{1+\ep}+\frac{4 \ep (8 \ep -55) \text{Sec}\left[\frac{2 \ep \pi }{1+\ep}\right]c_s^2}{(55+8 f_X+2 \ep (15 \ep +12 \ep f_X-47-16 f_X)) \Gamma\left[\frac{4}{1+\ep}-3\right]} +{\cal O} (c_s^4),
\ee
which is an exact result in $\ep$ (the solution to all orders in $c_s$ can be found in~\cite{nm}). Interestingly this means we have a generically blue running of non-Gaussianities (as long as $\epsilon \not< 0$), resulting in larger interactions and hence enlarged signals on smaller scales. In other words, primordial non-Gaussianities measured on e.g. galaxy cluster scales would be larger than those measured at CMB scales. However, this also means interactions will eventually become strongly coupled for sufficiently small scales.     
Following~\cite{baumann} we take the ratio of cubic and quadratic Lagrangians as our measure of strong coupling, requiring 
\be
\frac{{\cal L}_3}{{\cal L}_2} \sim {\cal O}(1,\epsilon,f_X)\frac{\zeta}{c_s^2} \ll 1
\ee
for fluctuations to be weakly coupled, roughly corresponding to $\fnl \ll 10^5$. If this condition breaks (at horizon crossing, where n-point correlation functions are evaluated here), quantum loop corrections are no longer suppressed and a perturbative treatment is no longer applicable. We now impose a minimal constraint of at least $\sim 10$ e-folds of weakly coupled inflation governed by action~\eqref{PX}, corresponding to the window of scales where primordial non-Gaussianity may be observable (from CMB, $k^{-1} \sim 10^3$ Mpc, to galactic scales, $k^{-1} \sim 1$ Mpc).\footnote{Beyond those 10 e-folds several options exist, depending on the UV-completion of the low-energy effective $P(X,\phi)$ theory~\cite{baumanngreen,gelaton}: new degrees of freedom may become important, resulting in an inflationary weakly coupled multi-field theory, the dispersion relation may change, a strongly coupled phase of inflation may take place,...} 

Depending on the size of $\fnl^{equi}$ at CMB scales, this results in different bounds on $n_{\rm NG}$ as shown in figure~\ref{fig2}. In terms of the scale $K$ the appropriate range here corresponds to $K_{\rm gal}/K_{\rm CMB} \simeq 10^3$. For $\fnl$ this means $f_{\rm NL}^{equi} ({\rm CMB})\ \approx \ 10^{-3(n_{\rm NG} - 1)} f_{\rm NL}^{equi} ({\rm Gal})$. If the bound on $n_{\rm NG}$ is satisfied, non-Gaussian interactions remain under perturbative control throughout the range of observable scales.
In the optimistic scenario with detectable CMB non-Gaussianities, i.e. $\fnl^{equi}({\rm CMB})  \gtrsim {\cal O}(10)$, we can combine these constraints with equation~\eqref{nng} to put an upper bound on $\epsilon$: $\epsilon \lesssim 0.3$.~\cite{kp,nm} If $\fnl^{equi}({\rm CMB})  \gtrsim {\cal O}(100)$ the bound is strengthened to $\epsilon \lesssim 0.2$. This shows how one can constrain the amount of slow-roll violation by requiring the action~\eqref{PX} to be a valid effective field theory over the observable range of scales for primordial fluctuations.\footnote{Note that eqn.~\eqref{nng} shows that these bounds receive ${\cal O}(c_s^2)$ corrections.~\cite{nm}}

\end{document}